\documentclass[12pt,letterpaper]{article}
\usepackage{setspace}
\usepackage[margin=1in]{geometry}
\usepackage[colorlinks,citecolor=blue]{hyperref}
\usepackage{mathrsfs}
\usepackage{bm, physics}
\usepackage{color}
\usepackage{caption}
\usepackage{natbib}

\usepackage{graphicx}
\usepackage{algorithm}
\renewcommand{\hat}{\widehat}
\renewcommand{\bar}{\overline}

\usepackage{algpseudocode}

\makeatletter
\def\maxwidth{ %
  \ifdim\Gin@nat@width>\linewidth
    \linewidth
  \else
    \Gin@nat@width
  \fi
}
\makeatother

\usepackage{color}
\definecolor{fgcolor}{rgb}{0, 0, 0}

\usepackage{framed}
\makeatletter
 {\par\unskip\endMakeFramed%
 \at@end@of@kframe}
\makeatother

\definecolor{shadecolor}{rgb}{.97, .97, .97}
\definecolor{messagecolor}{rgb}{0, 0, 0}
\definecolor{warningcolor}{rgb}{1, 0, 1}
\definecolor{errorcolor}{rgb}{1, 0, 0}
\usepackage{alltt}
\usepackage{natbib}
\usepackage{amsfonts,amssymb,graphics,epsfig,verbatim,bm}
\usepackage{latexsym,amsmath,url,amsbsy}
\usepackage{graphicx, cleveref}
\usepackage{booktabs,setspace}
\usepackage{caption}
\usepackage{multirow}

\usepackage{color}

\renewcommand{\hat}{\widehat}
\renewcommand{\bar}{\overline}

\begin{document}

\title{Heterogeneity Learning for SIRS model: an Application to the COVID-19}

 \author{Guanyu Hu~~~Junxian Geng}
\date{}
\maketitle 

\begin{abstract}
 We propose a Bayesian Heterogeneity Learning approach for Susceptible-Infected-Removal-Susceptible (SIRS) model that allows underlying clustering patterns for  transmission rate, recovery rate, and loss of immunity rate for the latest coronavirus (COVID-19) among different regions. Our proposed method provides simultaneously inference on parameter estimation and clustering information which contains both number of clusters and cluster configurations. Specifically, our key idea is to formulates the SIRS model into a hierarchical form and assign the Mixture of Finite mixtures priors for heterogeneity learning. The properties of the proposed models are examined and a Markov chain Monte Carlo sampling algorithm is used to sample from the posterior distribution. Extensive simulation studies are carried out to examine empirical performance of the proposed methods. We further apply the proposed methodology to analyze the state level COVID-19 data in U.S.
\bigskip

\noindent 
\textit{keywords: Bayesian Nonparametric, Cluster Learning, Infectious Diseases, MCMC, Mixture of Finite Mixtures} 
\end{abstract}

\maketitle
\newpage 
    % main body
\section{Introduction}\label{sec:intro}
The Coronavirus Disease 2019 (COVID-19) has created a profound public health emergency around world. It has become an epidemic with more than 5,000,000 confirmed infections worldwide as on May 21 2020. The spreading speed of COVID-19 which is caused by a new coronavirus is faster than severe acute respiratory syndrome (SARS) and Middle East respiratory syndrome (MERS). Recently, the risk of COVID-19 has been a significant public-health concern and people pay more attention on precise and timely estimates and predictions of COVID-19.  The Susceptible-Infectious-Recovered (SIR) model and its variation approaches, such as Susceptible-Infected-Removal-Susceptible (SIRS) \citep{kermack1932contributions,kermack1933contributions} and Susceptible-Exposed-Infected-Removal (SEIR) model \citep{hethcote2000mathematics}, have been widely discussed to study the dynamical evolution of an infectious disease in a certain region. There are rich literatures producing early results on COVID-19 based on SIR model and its variations \citep{wu2020nowcasting,read2020novel,tang2020estimation}. From statistician's perspectives, building a time-varying model under SIR and its variations is also fully discussed for COVID-19 \citep{chen2020time,sun2020discussion,jo2020analysis}. In most existing literature, people focus more on dynamic regimes of the SIR models for COVID-19. They lack discussions on heterogeneity pattern of COVID-19 among different regions.

The aim of this paper is to propose a new hierarchical SIRS model for detecting heterogeneity pattern of COVID-19 among different regions under a Bayesian framework. Bayesian nonparametric methods such as Dirichlet process (DP) offer choices to do simultaneously inference on parameters' estimation and parameters' heterogeneity information which contains the number of clusters and clustering configurations. Compared with existing approaches such as finite mixtures models, Bayesian nonparametric approach does not need to pre-specify the number of clusters, which provides probabilistic framework for simultaneous inference of the number of clusters and the clustering labels. \citet{miller2013simple} points out that the estimation of the number of clusters under Dirichlet process mixture (DPM) model is inconsistent which will produce extremely small clusters. One remedy for over-clustering problem under DPM is mixture of finite mixtures model (MFM) \citep{miller2018mixture}. The clustering properties of MFM are fully discussed in \citet{miller2018mixture,geng2019probabilistic} and it has been widely applied in different areas such as regional economics \citep{hu2020bayesian}, environmental science \citep{geng2019bayesian}, and social science \citep{geng2019probabilistic}. Thus, the key idea of this paper is to assign MFM priors on different parameters of the SIRS model to capture the heterogeneity of parameters among different regions. The contribution of this paper are two-fold. First, we formulate a Bayesian heterogeneity learning model for SIRS under MFM. To our best knowledge, this is the first time when MFM is applied into epidemiology models such as SIRS. Our proposed Bayesian approach successfully captures the heterogeneity of three different parameters under the SIRS model among different regions while also considering uncertainty in estimation of the number of clusters. Several interesting findings based on our proposed method are discovered for COVID-19 data in US.   

This paper is organized as follows. Section \ref{sec:example} presents the motivating data we analyze. We discuss our proposed Bayesian hierarchical model for heterogeneity learning under SIRS model framework in Section \ref{sec:method}. The performance of our proposed method is illustrated via simulation studies in Section \ref{sec:simu}. Section \ref{sec:real_data} is devoted to the analysis of state level COVID-19 data in U.S. A brief discussion is presented in Section \ref{sec:discussion}.

\section{Motivating Data}\label{sec:example}
Our motivating data comes from the COVID tracking project \url{https://covidtracking.com}.
State Level COVID-19 Data are recorded for the 50
states plus Washington, DC. For simplicity, we refer to them as ``51 states'' in the rest of this paper.  Up to June 10th, 2020, United States totally confirmed 2,043,031 cases. 114,533 people died because of COVID-19, and 607,279 people are recovered from COVID-19. The fatality rate of COVID-19 is $5.6\%$. 

\begin{figure}
    \centering
    \includegraphics[width=2in]{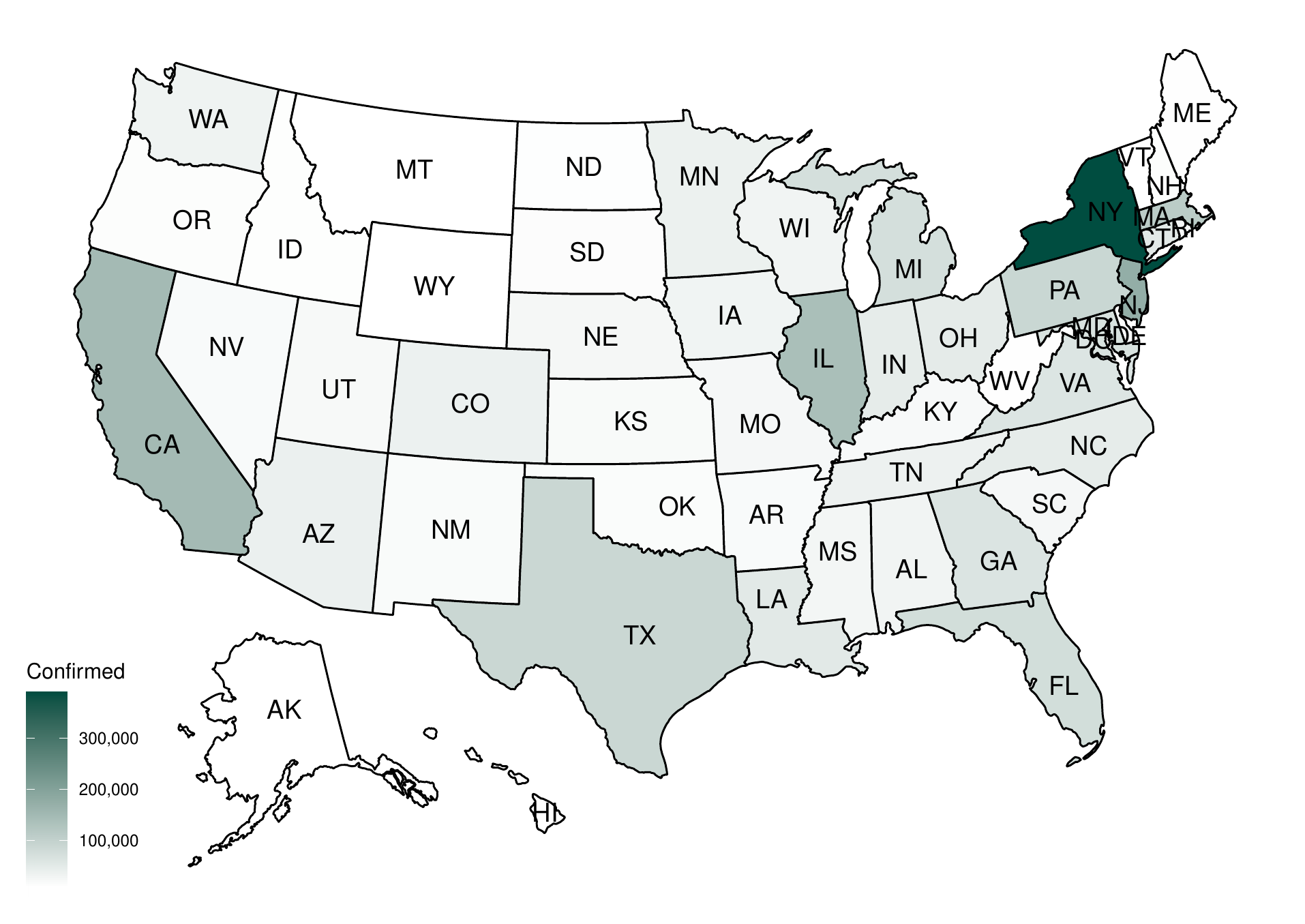}
    \includegraphics[width=2in]{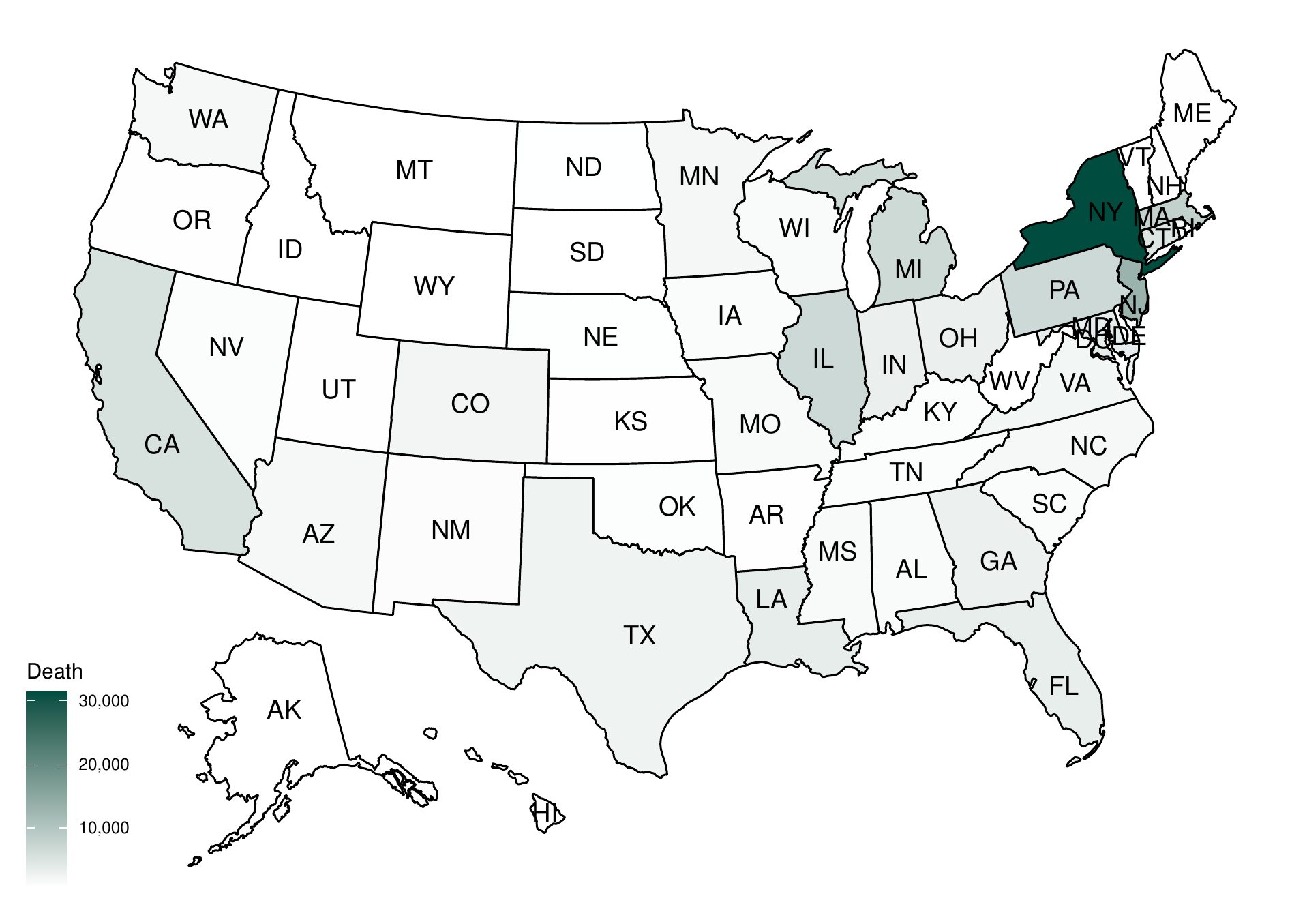}\\
    \includegraphics[width=2in]{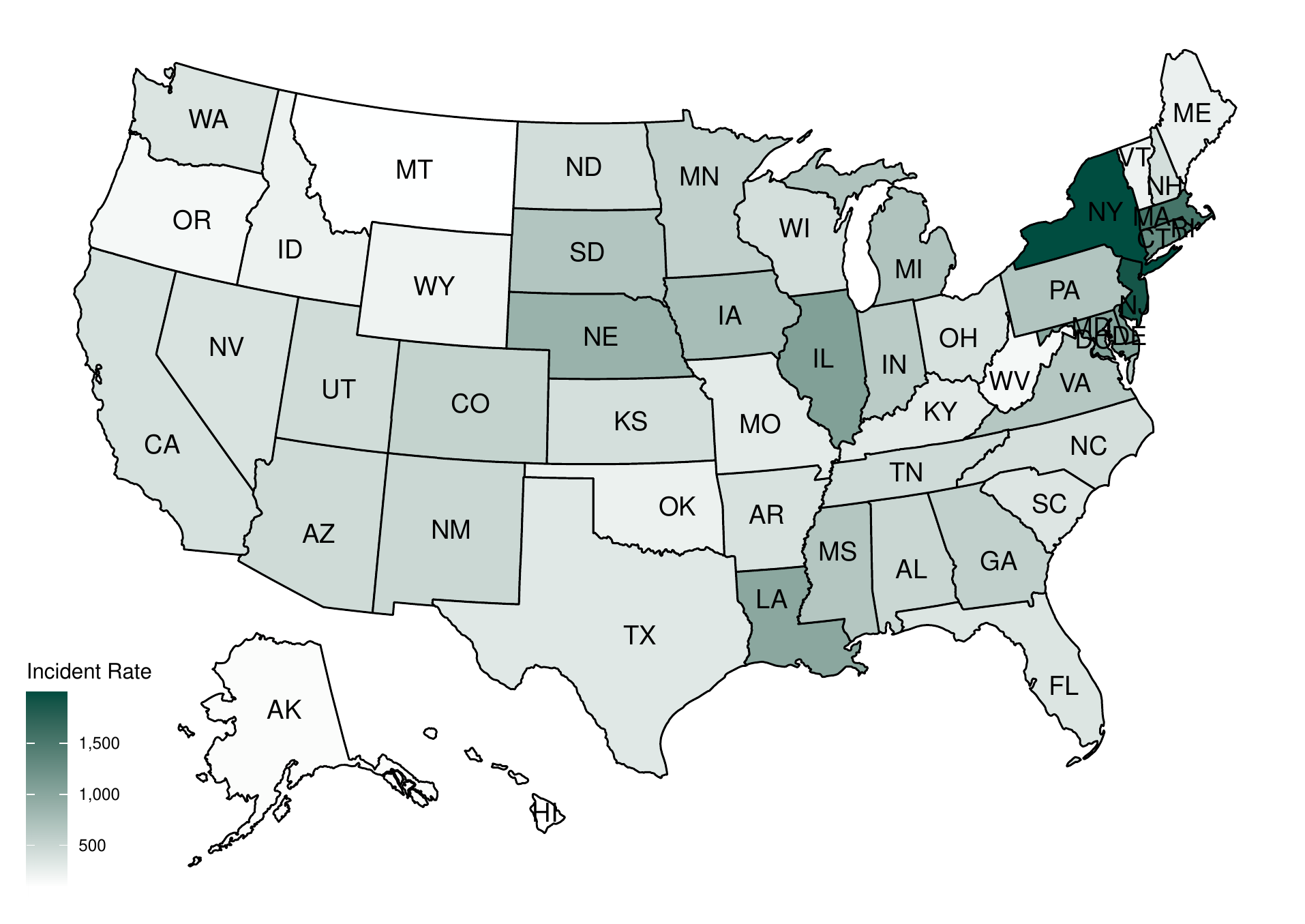}
    \includegraphics[width=2in]{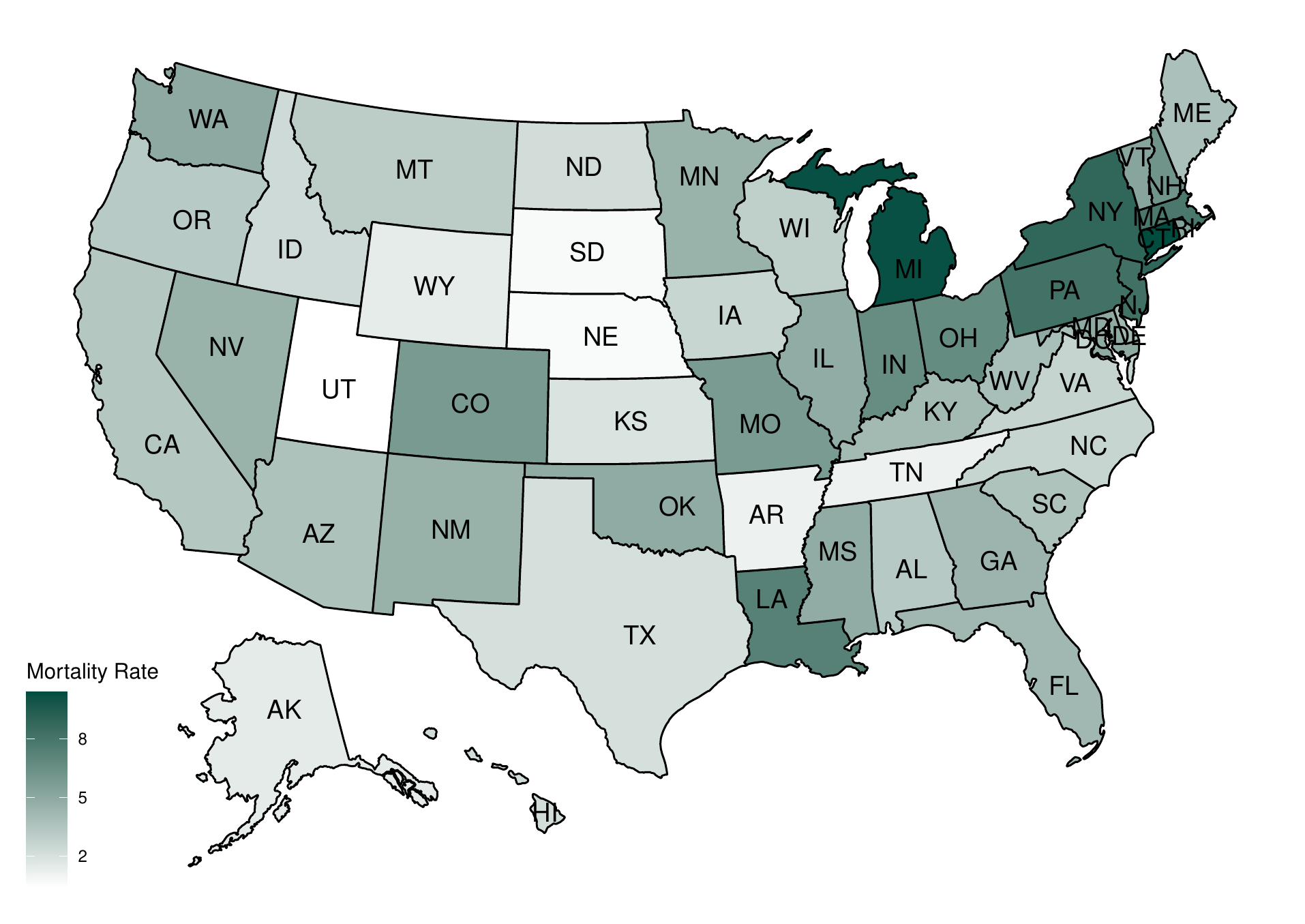}
    \caption{Exploratory Analysis of COVID-19 on June 10th}
    \label{fig:exp_data}
\end{figure}

Figure \ref{fig:exp_data} shows state level confirmed numbers, death numbers, incident rate, and mortality rate. We can see that New York state has the highest confirmed number, death numbers, and incident rate; Connecticut has the highest mortality rate among 51 states; Montana has the least confirmed number; Alaska has the least death number; the incident rate of Hawaii is lowest among 51 states; and Texas has the lowest mortality rate. 

\section{Method}\label{sec:method}
\subsection{SIRS Model}
Compartment epidemic models are widely used to study infectious disease which spreads through human populations across a large region. SIR model \citep{kermack1927contribution} has been universally discussed for analyzing the dynamical evolution of an infectious disease in a large population. SIR model is extended to SIRS model for imperfect immunity situation \citep{kermack1932contributions,kermack1933contributions}. For a given time $t$, a fixed population can be split into three compartments: $S(t)$, $I(t)$, and $R(t)$, which denotes the number of susceptibles,  the number of infectious, and the number of ``recovereds'' (which includes deaths), respectively.
The dynamical process of SIRS model can be written as following nonlinear ordinary differential equations of three given compartments
\begin{equation}
    \begin{split}
        &\frac{d S}{d t}=-\beta S I/N+\phi R,\\
        &\frac{d I}{d t}=\beta S I/N-\gamma I,\\
        &\frac{d R}{d t}=\gamma I-\phi R,
    \end{split}
    \label{eq:ode}
\end{equation}
where $\beta$ denotes the average rate of contact per unit time multiplied by
the probability of disease transmission per contact between a susceptible and
an infectious subject, $\gamma$ denotes 
the rate of “recovery” per unit time, which is the rate at which infectious individuals are removed from being infectious due 
to recovery (or death), and $\phi$ denotes the rate of loss of immunity of recovered individuals per unit time, which is the 
rate at which recovered individuals become susceptible again \citep{anderson1992infectious,zhuang2014bayesian}.
By adding the equations in \eqref{eq:ode}, we notice that
\begin{equation*}
 \frac{d S}{d t}+ \frac{d I}{d t}+ \frac{d R}{d t}=0. 
\end{equation*}
Thus, the model postulates a fixed total population without entry and exits of demographic type. For example, there are no births or deaths caused by other than the disease we study in a certain time. The SIRS model assume the sum of all three compartments together is constant within a short period of time such that 
\begin{equation}
    S(t)+I(t)+R(t)=N,
    \label{eq:constraint}
\end{equation}
where $N$ is a fixed total population. In cases with discrete time $t=1,\ldots,T$ (in units of days), we have 
\begin{equation}
\begin{split}
S(t+1)	=S(t)-\beta S(t)I(t)/N+\phi R(t), \\
I(t+1)	=I(t)+\beta S(t)I(t)/N-\gamma I(t),\\
R(t+1) = R(t)+\gamma I(t)- \phi R(t),   
\end{split}
\label{eq:dicrete_time}
\end{equation}
with the same constraints as \eqref{eq:constraint}.

Based on the models in \eqref{eq:dicrete_time} and \eqref{eq:constraint} and assmuptions in \citep{dukic2012tracking}, the data model of SIRS assumes conditional independent Poisson distributions evolving at discrete time points. For a given time $t=1,\ldots,T$, the data models are

\begin{equation}
    Z_R(t)|P_R(t)\sim \text{Possion} (N\times P_R(t)),
    \label{eq:poisson_rec}
\end{equation}
and  
\begin{equation}
    Z_I(t)|P_I(t)\sim \text{Possion} (N\times P_I(t)),
    \label{eq:poisson_inf}
\end{equation}
where $Z_R(t)$ and $Z_I(t)$
 are the observed number of ``recovereds'' (includes deaths) and infectious individuals at time $t$, respectively; $N$ is known total number of population and $Z_S(t)=N-Z_I(t)-Z_R(t)$; $P_R(t)$ and $P_I(t)$ are underlying true rates of recovered and infectious individuals. Thus, our observed data are $\{(Z_R(t),Z_I(t))\}:t=1,2\ldots,T$. Based on the relationship between the number of ``recovereds'', infectious, and suspects, we have
 \begin{equation}
     \label{eq:prop_quailance}
     P_R(t)+P_I(t)+P_S(t)=1,
 \end{equation}
 where $P_S(t)$ underlying rate of susceptible individuals. 
 
 Similar to \citep{zhuang2014bayesian}, we have following hidden processes:
 \begin{equation}
 \label{eq:evolution}
     \begin{split}
         P_R(t+1)&=P_R(t)+\gamma P_I(t)-\phi P_R(t),\\
         P_I(t+1)&=P_I(t)+\beta  P_S(t)P_I(t)-\gamma P_I(t),\\
         P_S(t+1)&=P_S(t)-\beta P_S(t)P_I(t)+\phi P_R(t).
     \end{split}
 \end{equation}

In order to model the hidden uncertainties in SIRS model, we define following transformation of $P_R(t)$, $P_I(t)$ and $P_S(t)$ based on \eqref{eq:prop_quailance}
\begin{equation}
    \label{eq:transformation}
    \begin{split}
        W_S(t)&\equiv \log(\frac{P_S(t)}{P_R(t)}),\\
        W_I(t)&\equiv \log(\frac{P_I(t)}{P_R(t)}).\\
    \end{split}
\end{equation}
The time-varying process of $W_R(t)$ and $ W_I(t)$ is defined as 
\begin{equation}
    \label{eq:time_varying_process}
    \begin{split}
        W_S(t+1)&=\mu_S(t)+\epsilon_S(t+1),\\
        W_I(t+1)&=\mu_I(t)+\epsilon_I(t+1),
    \end{split}
\end{equation}
where $\epsilon_S(t)\sim N(0,\sigma^2_S)$ and $\epsilon_I(t)\sim N(0,\sigma^2_I)$ for $t=1,2,\ldots,T$. Based on \eqref{eq:prop_quailance} and \eqref{eq:evolution}, we have 
\begin{equation}
\label{eq:mu_S}
\begin{split}
     \mu_S(t)=&W_S(t)\\
     &+\log (1+\frac{\phi}{\exp(W_S(t))}-\frac{\beta  \exp(W_I(t))}{1+ \exp(W_S(t))+ \exp(W_I(t))})\\
     &+\log (\frac{1}{1+\gamma  \exp(W_I(t))-\phi})    
\end{split}
\end{equation}
and 
\begin{equation}
\label{eq:mu_I}
\begin{split}
     \mu_I(t)=& W_I(t)\\
         &+\log(1-\gamma+\frac{\beta  \exp(W_S(t))}{1+\exp(W_S(t))+\exp(W_I(t))})\\
         &+\log(\frac{1}{1+\gamma \exp(W_I(t))-\phi})   
\end{split}
\end{equation}
Based on the transformation in \eqref{eq:transformation}, we have our data in \eqref{eq:poisson_rec} and \eqref{eq:poisson_inf} as 

\begin{equation}
    \begin{split}
        Z_R(t)|W_S(t),W_I(t)\sim \text{Possion} \left(N\times \frac{1}{1+\exp(W_S(t))+\exp(W_I(t))}\right),\\
        Z_I(t)|W_S(t),W_I(t)\sim \text{Possion} \left(N\times \frac{\exp(W_I(t))}{1+\exp(W_S(t))+\exp(W_I(t))}\right).
    \end{split}
    \label{eq:final_sirs}
\end{equation}
For the simplicity, we refer the model from \eqref{eq:time_varying_process} to \eqref{eq:final_sirs} as $\{(Z_R(t),Z_I(t),N),t=1,2\ldots,T\}\sim \text{SIRS}(\beta,\gamma,\phi,\sigma^2_S,\sigma^2_I)$.
Based on the transmission rate and recover rate, the basic reproduction number, $R_0$, can be calculated by 
\begin{equation}
\label{eq:reproduce}
    R_0=\frac{\beta}{\gamma}.
\end{equation}

\subsection{Heterogeneity Learning}
In section \ref{sec:example}, our motivating data is at state level in US and we are interested in whether there are heterogeneity patterns on the parameters of interest among different states. As an assumption, we believe that different states might have different parameters, however, some states will share similar pattern in transmission rate, recovery rate, or loss of immunity rate. Next, we introduce nonparametric Bayesian methods for heterogeneity learning of SIRS parameters over $n$ different regions. In this section, we focus on the the transmission rate $\beta$ for different regions. Recovery rate and loss of immunity rate can be parameterized in the same way. 

Let $z_1, \ldots, z_n\in \{1,\ldots,k\}$ denote clustering labels of $n$ regions and
$\beta_1,\ldots,\beta_n$ denote the corresponding parameters in SIRS model for $n$ regions.
Our goal is to cluster $\beta_1,\ldots,\beta_n$ into $k$ clusters with distinct values $\beta^*_1,\ldots,\beta^*_k$, which is usually unknown in practice.
A popular solution for unknown $k$ is to introduce the Dirichlet process mixture prior models
\citep{antoniak1974mixtures} as following: 
\begin{eqnarray}\label{eq:DPMM}
\beta_i \sim  G,
\quad G \sim  DP(\alpha G_0),
\end{eqnarray}
where $G_0$ is a base measure and $\alpha$ is a concentration parameter. 
If a set of values of $\beta_1,\ldots,\beta_n$
are drawn
from $G$, a conditional prior can be obtained by integration
\citep{blackwell1973ferguson}:
\begin{eqnarray}\label{eq:DPMM1}
p(\beta_{n+1}\mid \beta_1,\ldots,\beta_n) =
\dfrac{1}{n+\alpha}\sum_{j=1}^n\delta_{\beta_j}(\beta_{n+1}) +
\dfrac{\alpha}{n+\alpha}G_0(\beta_{n+1}).
\end{eqnarray}
Here, $\delta_{\beta_j}(\beta_{\ell}) = I(\beta_\ell =
\beta_j)$ is a point mass at $\beta_j$. We can obtain the following equivalent models by introducing cluster membership $z_j$'s and letting the unknown number of clusters $k$ go to infinity \citep{neal2000markov}. 
\begin{equation}\label{eq:DPMM2}
\begin{split}
z_i \mid  \bm{\pi} & \sim \text{Discrete} (\pi_1,\ldots,\pi_k), \\
\beta^*_c & \sim G_0\\
\bm{\pi} & \sim \text{Dirichlet}(\alpha/k,\ldots ,\alpha/k) 
\end{split}
\end{equation}
where $\bm{\pi}=(\pi_1,\ldots,\pi_k)$. 
%By integrating out mixing proportions $\bm{\pi}$, 
%we can obtain the prior
%distribution of $(z_1, z_2, \ldots, z_n)$ that allows automatic inference on the
%number of clusters $k$, which is also well known as the Chinese restaurant
%process
%\citep
%[CRP;][]{aldous1985exchangeability,pitman1995exchangeable,neal2000markov}.
%The CRP is defined through the following conditional distribution \citep[P\'{o}lya urn
%scheme,][]{blackwell1973ferguson}:
%\begin{eqnarray}\label{eq:crp}
%P(z_{i} = c \mid z_{1}, \ldots, z_{i-1})  \propto   
%begin{cases}
%\abs{c}  , &  \text{at an existing table labeled}\, c\\
%\alpha,  & \text{if} \, $c$\,\text{is a new table}
%\end{cases},
%\end{eqnarray}
%where $\abs{c}$ is the size of cluster $c$. 
In addition, the distribution of $z_i$ can be marginally given by a stick-breaking representation
\citep{sethuraman1994constructive} of Dirichlet process (DP) as 
\begin{equation}
	\begin{split}
		z_i&\sim \sum_{h=1}^\infty \pi_h \delta_h,\\
		\pi_h&=\nu_h\prod_{\ell\leq h}(1-\nu_\ell),\\
		\nu_h&\sim \text{Beta}(1,\alpha),
	\end{split}
	\label{eq:s-b_construct}
\end{equation}
where $\delta_h$ is the Dirac function with mass at $h$. 

However, \cite{miller2018mixture} proved that the DP mixture model
produces extraneous clusters in the posterior leading to inconsistent
estimation of the {\em number of clusters} even when the sample size grows to infinity.
A modification of DP mixture model called Mixture of finite mixtures (MFM) model is
proposed to circumvent this issue \citep{miller2018mixture}:
\begin{eqnarray}\label{eq:MFM}
k \sim p(\cdot), \quad (\pi_1, \ldots, \pi_k) \mid k \sim \mbox{Dirichlet}
(\eta,
\ldots, \eta), \quad z_i \mid k, \bm{\pi} \sim \sum_{h=1}^k  \pi_h
\delta_h,\quad
i=1, \ldots, n, 
\end{eqnarray}
where~$p(\cdot)$ is a proper probability mass function (p.m.f.) on~$\{1, 2,
\ldots, \}$.

Like the stick-breaking representation in \eqref{eq:s-b_construct} of Dirichlet process, the MFM also has a similar construction. If we choose $k-1 \sim \mbox{Poisson}(\lambda)$ and
$\eta=1$ in \eqref{eq:MFM}, the mixture weights~$\pi_1,\cdots,\pi_k$ is
constructed as follows:
\begin{enumerate}
	\item Generate $\eta_1,\eta_2,\cdots \overset{\text{iid}}{\sim}
\text{Exp}(\lambda)$,
	\item $k=\min\{j:\sum_{i=1}^j \eta_i\geq 1\}$,
	\item $\pi_i=\eta_i$, for $i=1,\cdots,k-1$,
	\item $\pi_k=1-\sum_{i}^{k-1}\pi_i$.
\end{enumerate}
For ease of exposition, we refer the stick-breaking representation of MFM above as $\text{MFM}(\lambda)$ with default choice of $p(\cdot)$ being $\mbox{Poisson}(\lambda)$ and $\eta=1$.

\subsection{Hierarchical Model}\label{sec:hierarchical}
In order to allow for simultaneously heterogeneity learning of three parameters
in SIRS model, the MFM prior is introduced for parameters
$\beta$, $\gamma$ and $\phi$ in the SIRS model. Our observed data are $\{(Z_R(t,\bm{s}_i),Z_I(t,\bm{s}_i),N_i): t=1,2,\ldots,T, i=1,2,\ldots,n\}$, where $t$ denotes each discrete time point and $i$ denotes each state. The hierarchical SIRS model with MFM is given as
\begin{equation}
    \begin{split}
     &\{(Z_R(t,\bm{s}_i),Z_I(t,\bm{s}_i),N_i),t=1,2\ldots,T\}\sim \text{SIRS}(\beta_{z^\beta_i},\gamma_{z^\gamma_i},\phi_{z^\phi_i},\sigma^2_{S,i},\sigma^2_{I,i}), i=1,2,\ldots,n\\
     &z^\beta_i\sim \text{MFM}(\lambda_\beta),i=1,2,\ldots,n,\\
     &z^\gamma_i\sim \text{MFM}(\lambda_\gamma),i=1,2,\ldots,n,\\
     &z^\phi_i\sim \text{MFM}(\lambda_\phi),i=1,2,\ldots,n,\\
     &\beta_{z^\beta_i}\sim G_\beta,\\
     &\gamma_{z^\gamma_i}\sim G_\gamma,\\
     &\phi_{z^\phi_i}\sim G_\phi,\\
     &\sigma^2_{S,i},\sigma^2_{I,i} \sim \text{IG}(0.01,0.01), i=1,2,\ldots,n,
    \end{split}
    \label{eq:hSIR_MFM}
\end{equation}
where $z^\beta$, $z^\gamma$, and $z^\phi$ denote the cluster assignments of parameter $\beta$,$\gamma$, and $\phi$, respectively. $G_\beta$, $G_\gamma$, and $G_\phi$ is the base distribution for parameter $\beta$,$\gamma$, and $\phi$, respectively. The choices of $G_\beta$, $G_\gamma$, and $G_\phi$ will be discussed in Section \ref{sec:prior}.

\subsection{Prior and Posterior}\label{sec:prior}
For the hierarchical SIRS model with MFM introduced in Section \ref{sec:hierarchical}, the set of parameters is denoted as $\Theta=\{\beta_{z^\beta_i}, \gamma_{z^\gamma_i},\phi_{z^\phi_i},\sigma^2_{S,i},\sigma^2_{I,i},\lambda_\beta,\lambda_\gamma,\lambda_\phi: i=1,2\ldots,n \}$. To complete the model, we now specify the joint prior distribution for the parameters. Based on the natural constraints generated by \eqref{eq:dicrete_time}, we have following distribution for bases distribution  $G_\beta$, $G_\gamma$ and $G_\phi$, respectively:
\begin{equation}
\begin{split}
       &\beta_{z_i^\beta}\sim \text{Uniform}(0,1) ,\\
     &\gamma_{z^\gamma_i}\sim \text{Uniform}(0,1),\\
     &\phi_{z^\phi_i}\sim \text{Uniform}(0,1).
\end{split}
\label{eq:base_distribution}
\end{equation}
For the hyperparameters for three MFM processes, we assign gamma prior $\text{Gamma}(1,1)$ on $\lambda_\beta,\lambda_\gamma,\lambda_\phi$.
With the joint prior distributions $\pi(\Theta)$, the posterior distribution of these parameters based on the
data $D=\{(Z_R(t,\bm{s}_i),Z_I(t,\bm{s}_i),N_i): t=1,2,\ldots,T, i=1,2,\ldots,n\}$ is given as 
\begin{equation}
    \pi(\Theta|(Z_R(t,\bm{s}_i),Z_I(t,\bm{s}_i),N_i): t=1,2,\ldots,T, i=1,2,\ldots,n)\propto L(D|\Theta)\times \pi(\Theta),
\end{equation}
where $L(D|\Theta)$ is the full data likelihood given from the model \eqref{eq:time_varying_process} to \eqref{eq:final_sirs}. The analytical form of the posterior distribution of $\pi(\Theta|(Z_R(t,\bm{s}_i),Z_I(t,\bm{s}_i),N_i): t=1,2,\ldots,T, i=1,2,\ldots,n)$ is unavailable. Therefore, we carry out the posterior inference using the MCMC sampling algorithm to sample from the posterior distribution and then obtain the posterior estimates of the unknown parameters. Computation is facilitated by the \textbf{nimble} package \citep{de2017programming} in R which generates \textbf{C++} code for faster computation.

\subsection{Group Inference via MCMC Samples}\label{sec:dalh}
After obtaining posterior samples, an important task is do inference on posterior samples.
Using posterior mean or posterior median for grouping label $\bm{z}$ is not suitable. Instead, inference on the clustering configurations is obtained employing the modal clustering method of
\citep{dahl2006model}. The inference is based on the membership matrices of posterior samples,
$B^{(1)},\ldots,B^{(M)}$, where $B^{(t)}$ for the $t$-th post-burn-in MCMC iteration is defined as:
\begin{align}
B^{(t)} = [B^{(t)}(i,j)]_{i,j\in \{1:n\}} = 1(z_i^{(t)} = z_j^{(t)})_{n\times n},~~
t=1,\ldots,M,
\end{align}
Here $1(\cdot)$ denotes the indicator function, which means
% with $B^{(t)}(i,j) \in \{0,1\}$ for all $i,j = 1,...,n$. 
$B^{(t)}(i,j)=1$ indicates observations $i$ and~$j$ are in the same cluster in
the $t$-th posterior sample post burn-in. After obtaining the membership matrices of the posterior samples, a Euclidean mean for membership matrices is
calculated by:
\begin{equation*}
  \bar{B} =\frac{1}{M} \sum_{t=1}^M B^{(t)}.
\end{equation*}
Based on $\bar{B}$ and $B^{(1)},\ldots,B^{(M)}$, we find the iteration with the least squares distance to $\bar{B}$ as
\begin{align}
C_{L} = \text{argmin}_{t \in (1:M)} \sum_{i=1}^n \sum_{j=1}^n \{B(i,j)^{(t)} -
\bar{B}(i,j)\}^2.
\end{align}  
The estimated parameters, together with the cluster assignments $\bm{z}$, are
then extracted from the $C_{L}$-th post burn-in iteration.

\section{Simulation}\label{sec:simu}

In this section, we investigate the performance of the hierarchical SIRS model with MFM from a variety of measures.

\subsection{Simulation Settings and Evaluation Metrics}
\begin{figure}
    \centering
    \includegraphics[width=3 in]{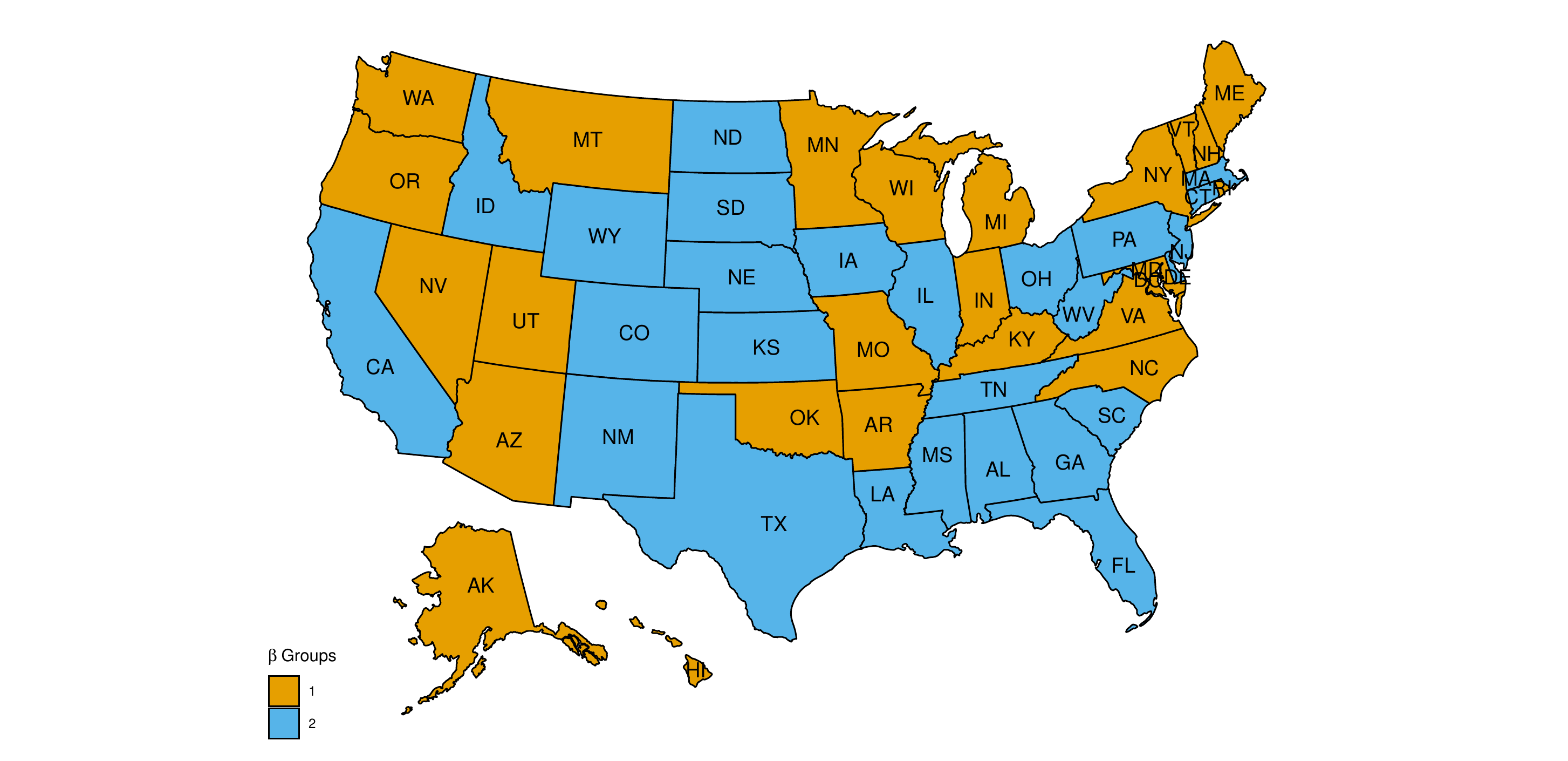}
    \includegraphics[width=3 in]{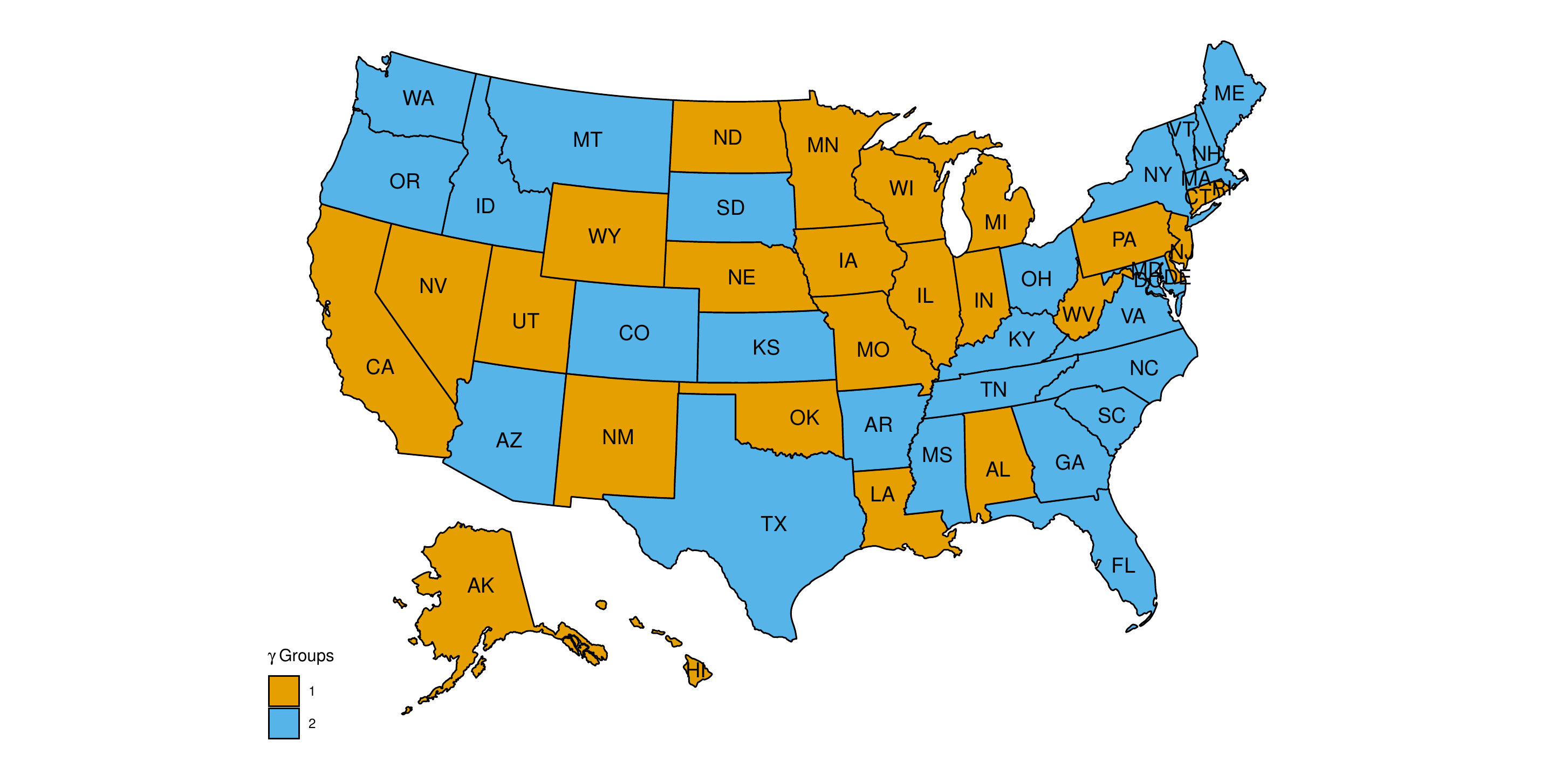}
    \includegraphics[width=3 in]{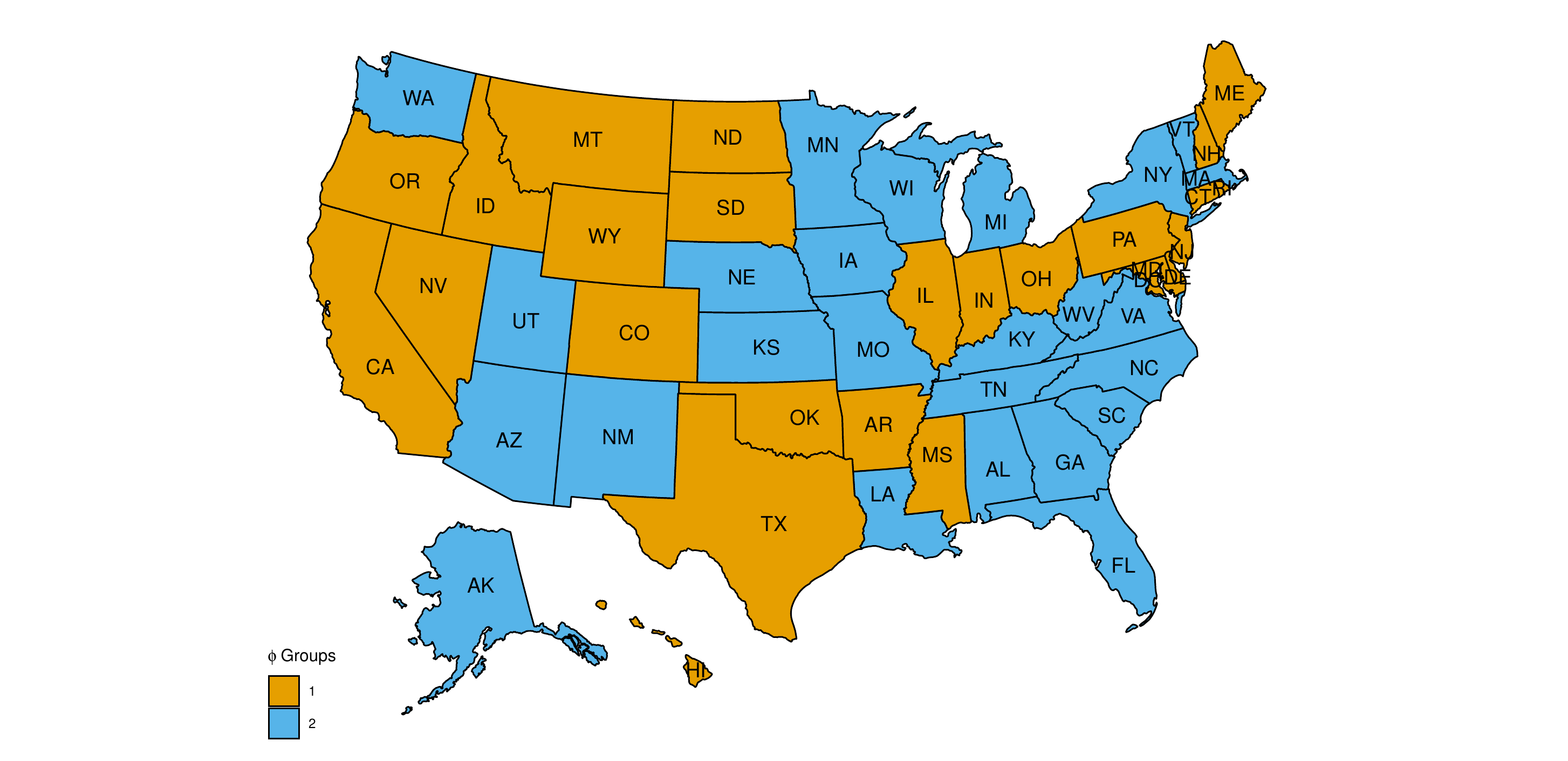}
    \caption{Grouping Labels for $\beta$, $\gamma$, and $\phi$}
    \label{fig:simu_group}
\end{figure}

In order to mimic the real dataset we analyze, we choose $n=51$ and the population for each location is assigned as the real data population for $51$ states. The time length $T$ equals 30 for all the simulation replicates. The total number of replicates in our simulation study is 100. For each parameter, we have two different groups and we set the true values of the parameters $\beta_1=0.06,\beta_2=0.6$, $\phi_1=0.06,\phi_2=0.6,$, and $\gamma_1=0.06,\gamma_2=0.6$. We randomly assign the labels to $51$ locations and fix them over 100 replicates. The grouping labels for three parameters is given in Figure \ref{fig:simu_group}.

For each replicates, we have $15,000$ iterations MCMC samples and have first $5,000$ iterations burn-in in order to obtain samples from every 5th iteration of the last $10,000$ iterations.

The performance of the posterior estimates of parameters were evaluated by the mean bias (MB) and the mean standard deviation (MSD)
in the following ways, take $\beta$ as an example:
\begin{gather*}
\text{MAB} = \frac{1}{100} \sum_{r=1}^{100} \bigg{\{}\frac{1}{n} \sum_{i=1}^{n}
\hat{\beta}^{r}(\bm{s}_i) - \beta(\bm{s}_i) \bigg{\}}, \\
\text{MSD} = 
\sqrt{\frac{1}{100}
	\sum_{r=1}^{100} \bigg{\{}\frac{1}{n} \sum_{i=1}^{n}\left(\hat{\beta}^{r}(\bm{s}_i) - \bar{\hat{\beta}}(\bm{s}_i)
	\right)^2}\bigg{\}},
%\\
%\text{MMSE} =  \frac{1}{100}
%\sum_{r=1}^{100}
%\frac{1}{n} \sum_{i=1}^{n} \left(\hat{\beta}^{r}(\bm{s}_i) - \beta(\bm{s}_i) \right)^2,
%\text{MCR} =  \frac{1}{100}
%\sum_{r=1}^{100} \frac{1}{N} \sum_{i=1}^{N}
%1 \left( \hat{\theta}_{i,r} \in \text{95\% HPD interval}
%\right),
\end{gather*}
where $\bar{\hat{\beta}}(\bm{s}_i)$ is the mean of the posterior estimate over 100 replicates.

For clustering estimation evaluation, the estimated number of clusters $\hat{K}$ for each replicate is summarized from the MCMC iteration picked by Dahl's method. Rand Index \citep[RI;][]{rand1971objective} is applied to evaluate cluster configuration. The RI is calculated by R-package \textbf{fossil} \citep{vavrek2011fossil}. A higher value of the RI represents higher accuracy of clustering The average RI (MRI) was calculated as the mean of RIs over the 100 replicates.

\subsection{Simulation Results}
The parameter estimation performance and clustering performance results are shown in Table \ref{tab:simu_estimate} and 
Table \ref{tab:group_estimate}.
\begin{table}[h!]
	\centering
	\caption{Estimation Performance for Simulation Study }\label{tab:simu_estimate}
	\begin{tabular}{ccc}
		\toprule
	 Parameter& MB & MSD  \\
		\midrule 
		 $\beta_1$ &  0.008 & 0.021   \\
		 $\beta_2$& -0.072 &0.152\\
         $\gamma_1$ &  0.007 & 0.017   \\
          $\gamma_2$ &  -0.068 & 0.151    \\
        $\phi_1$ &  0.012 & 0.023    \\
         $\phi_2$ &  -0.069 & 0.149    \\
		\bottomrule
	\end{tabular}
\end{table} 

\begin{table}[h!]
	\centering
	\caption{Grouping Performance for Simulation Study }\label{tab:group_estimate}
	\begin{tabular}{ccccc}
		\toprule
	 Parameter& MRI & S.D of RI & $\hat{K}$& S.D. of $\hat{K}$  \\
		\midrule 
		 $\beta$ &  0.854 & 0.058 &2.12& 0.33 \\
		 $\gamma$ &  0.857 & 0.057 &2.33 &0.55 \\
         $\phi$ &  0.847 & 0.059 & 2.31&0.54  \\

		\bottomrule
	\end{tabular}
\end{table} 

From the results shown in Table \ref{tab:simu_estimate}, we see that the MABs and MSDs of the parameters are both within a reasonable range. In general, performance of posterior estimates achieve a good target.

And we see that our proposed methods successfully recover the number of groups and grouping labels within a reasonable range for all three parameters from Table \ref{tab:group_estimate}.  Average rand index for all parameters around 0.85 indicate our proposed method truly recovers the group labels for all three parameters.  The mean of the estimated number of groups is close to true number of groups over 100 replicates.

\section{Real Data Analysis}\label{sec:real_data}

\subsection{30-Day Analysis from April 1st}
We analyze 30-Day data from April 1st, 2020. The reason why we analyze this time period data is that U.S. Government announced the suspension of entry as immigrants and nonimmigrants of certain additional persons who pose a risk of transmitting corona-virus \url{https://www.whitehouse.gov/presidential-actions/} on March 11th, 2020. From the April 1st, we can assume that there are very limited imported cases from outside U.S.. We analyze 30-day data based on the model in \eqref{eq:hSIR_MFM} and use the priors discussed in Section \ref{sec:prior}. We run $50,000$ MCMC iterations and burnin
the first $20,000$ iterations in order to obtain samples from every 10th iteration of the
last $30,000$ iterations. The group labels are obtained by Dalh's method in Section \ref{sec:dalh}.

\begin{figure}
    \centering
    \includegraphics[width=4in]{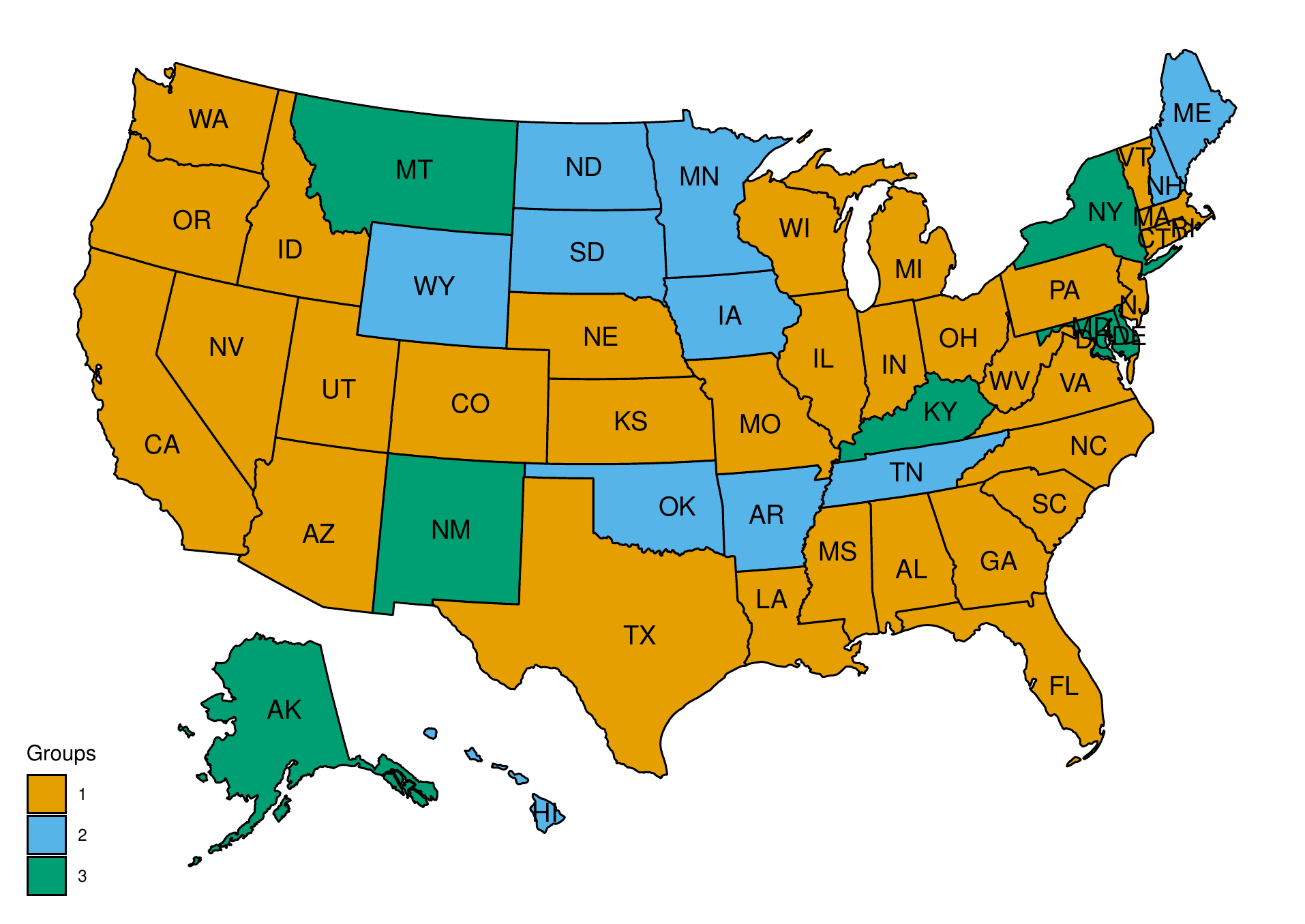}
    \caption{Group Labels for $\gamma$s of April 1st data}
    \label{fig:gamma_group}
\end{figure}

For $\beta$, one group is identified. $\beta=0.079$ with $95\%$ Highest Probability Density (HPD) interval \citep{chen1999monte} $(0.058, 0.098)$.
For $\gamma$, three groups are identified with  $\gamma_1=0.0054$ with HPD interval $(0.0021, 0.0207)$, $\gamma_2=0.0419$ with HPD interval $(0.0022, 0.0609)$ and $\gamma_3=0.0164$ with HPD interval$ (0.0035, 0.0241)$. Thirty three states including Alabama, Arizona, California, Colorado, Connecticut, District of Columbia, Florida, Georgia, Idaho, Illinois, Indiana, Kansas, Louisiana, Massachusetts, Michigan, Mississippi, Missouri, Nebraska, Nevada, New Jersey, North Carolina, Ohio, Oregon, Pennsylvania, Rhode Island, South Carolina, Texas, Utah, Vermont, Virginia, Washington, West Virginia, Wisconsin, belong to group one. Eleven states including Arkansas, Hawaii, Iowa, Maine, Minnesota, New Hampshire, North Dakota, Oklahoma, South Dakota, Tennessee, Wyoming , belong to group 2. And seven states including Alaska, Delaware, Kentucky, Maryland, Montana, New Mexico, New York, belong to group 3. 
The grouping labels for $\gamma$s are shown in Figure \ref{fig:gamma_group}.

For $\phi$, one group is identified. $\phi=0.0015$ with HPD interval $(1.181\times10^{-7},0.0047)$.
% For $\phi$, three groups are identified. $\phi_= 0.5123$ with HPD interval $(0.2489, 0.9429)$, $\phi_2=0.6371$ with HPD interval $(0.3031, 0.9996)$, and $\phi_3=0.0008$ with HPD interval $(2.467\times 10^{-7}, 0.0026)$. Only one state, Nebraska, belongs group 1. And one state, Georgia, belongs to group 2. These two states have comparable higher rate of loss of immunity of recovered individuals than other 49 states. 

\begin{figure}
    \centering
    \includegraphics[width= 4in]{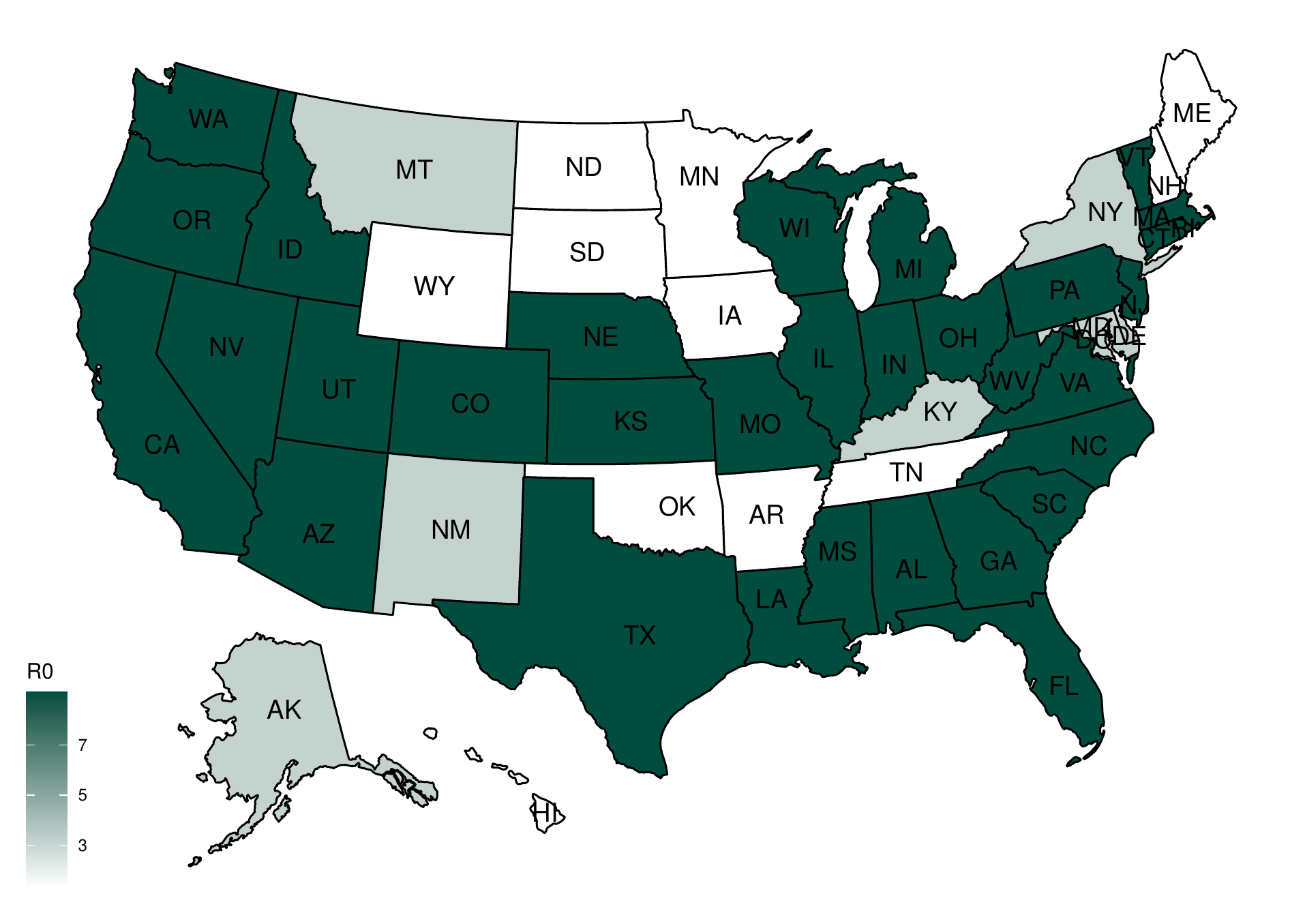}
    \caption{$R_0$ for 51 States from April 1st}
    \label{fig:R0}
\end{figure}

With the estimated values of $\beta$ and $\gamma$, the basic reproduction number, $R_0$, is calculated among different states. The values of $R_0$ among different states are shown in Figure \ref{fig:R0}.

\subsection{30-Day Analysis from May 1st}
The second time period we analyze is from May 1st, 2020. Other settings are same with previous analysis. 

For $\beta$, one group is identified. $\beta=0.0042$ with $95\%$ Highest Probability Density (HPD) interval \citep{chen1999monte} $(3.056\times 10^{-8}, 0.1083)$. Compared with previous 30-day data, in this time period, the transmission rate decrese a lot.
For $\gamma$, two groups are identified with  $\gamma_1=0.0381$ with HPD interval $(0.0048, 0.3713)$ and $\gamma_2=0.0007$ with HPD interval $( 0.0003, 0.0013)$. Two states, Oregon and Vermont, are identified in group 1. Other states are identified in group 2. For $\phi$, one groups is identified. $\phi=0.0006$ with HPD interval $(2.747\times 10^{-7}, 0.0026)$.

With the estimated values of $\beta$ and $\gamma$, the basic reproduction number, $R_0$, is calculated among different states. There are two different groups for the basic reproduction number. One group include Oregon and Vermont with $R_0=0.1102$. The other group includes other 49 states with $R_0=5.4619$. Comparing to the 30 days period from April 1st, we can see a decrease for $R_0$ in general.

\section{Discussion}\label{sec:discussion}
In this paper, we develop a nonparametric Bayesian heterogeneity learning method for SIRS model based on Mixture of Finite Mixtures
model. This statistical framework was motivated by the heterogeneity of COVID-19 pattern among different regions.
Our simulation results indicate that the proposed method can recover
the heterogeneity pattern of parameters among different regions. 
Illustrated by the analysis of COVID-19 data in U.S., our proposed methods reveal the heterogeneity pattern among different states.

In addition, three topics beyond the scope of this paper are worth
further investigation. First, we can add spatially dependent structure \citep{hu2020bayesian,zhao2020bayesian}
on the heterogeneity of different states. Second, our model assumes parameters are constant over a certain time period. 
Building heterogeneity learning model with time varying coefficients is an interesting future work. Finally, proposing a 
measurement error correction for SIRS devotes another interesting future work.

\bibliography{main.bib}
\bibliographystyle{chicago}
\end{document}